\documentclass[twocolumn,letter]{jpsj2}
\def\gsim{\mathop {\vtop {\ialign {##\crcr 
$\hfil \displaystyle {>}\hfil $\crcr \noalign {\kern1pt \nointerlineskip
 } 
$\,\sim$ \crcr \noalign {\kern1pt}}}}\limits}
\def\lsim{\mathop {\vtop {\ialign {##\crcr 
$\hfil \displaystyle {<}\hfil $\crcr \noalign {\kern1pt \nointerlineskip
 } 
$\,\,\sim$ \crcr \noalign {\kern1pt}}}}\limits}

\def\runauthor{
}
\title{ A Poor Man's Derivation of Quantum Compass-Heisenberg Interaction:\\
Superexchange Interaction in J-J Coupling Scheme }
          
\author{ Hiroyasu {\sc Matsuura} and Masao {\sc Ogata} 
}
\inst
{
Department of Physics, University of Tokyo, 7-3-1 Hongo, Bunkyo, Tokyo 113-0033, Japan\\
}

\recdate
{
\today
}

\abst
{The exchange interaction between 5d electrons in t$_{2g}$ orbitals is derived in the J-J coupling scheme which is the appropriate basis in the case of strong spin-orbit interaction.
From simple calculations, it is found that ferromagnetic Ising interaction (quantum compass model) occurs due to a selection rule of hybridization between $\Gamma_7$ and $\Gamma_8$ orbitals, while antiferromagnetic Heisenberg interaction appears from hybridization between $\Gamma_7$ orbitals.
It is also found that the ferromagnetic Ising interaction becomes small as the spin-orbit interaction increases. 
Thus, the sign of exchange interaction changes from ferromagnetic Ising to antiferromagnetic Heisenberg interaction as the spin-orbit interaction increases. }

\kword{(5$d)^5$ electron, Large spin-orbit interaction, J-J coupling, Slater-Condon parameter, Quantum compass model 
}
\begin{document}
\sloppy

\maketitle
5d electron systems have attracted much interest since they show an unconventional metal-insulator transition induced by spin-orbit (SO) interaction~\cite{Kim}.In the insulating state, exchange interactions between 5d electrons play important roles to determine the electronic states.
In particular, since the SO interaction of 5d electron systems is much larger than that of 3d or 4d systems, we expect that anisotropy and sign of the exchange interaction are drastically different from those in 3d or 4d systems~\cite{Matsuura}.   

\begin{figure}[h]
\begin{center}
\rotatebox{0}{\includegraphics[angle=-90,width=0.8 \linewidth]{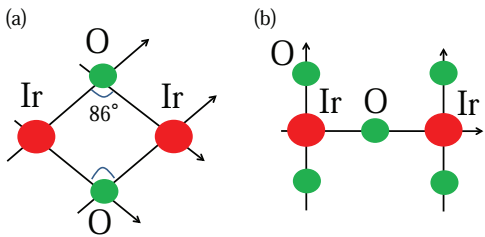}}
\caption{Schematic picture of (a) edge-shared bond and (b) corner-shared bond between Ir ions (red circles) and oxygen ions (green circles). In CaIrO$_3$, the Ir-O-Ir bond angle is about 86$^{\circ}$ as shown in Fig.~1(a). }
\label{Fig1}
\end{center}
\end{figure}

As a typical system of 5d electrons, CaIrO$_3$ has been extensively investigated.
Actually, CaIrO$_3$ has an insulating behavior at a room temperature and becomes a canted antiferromagnetic state below $T_N =115$K. 
Since the nominal valence of Ir is 4+ and there are six oxygens around an Ir ion, five 5d electrons occupy t$_{2g}$ orbitals on the Ir ion~\cite{Ohgushi1}.
Depending on the crystal structure, there are two kinds of bond geometries of Ir ions and surrounding oxygen ions: an edge-shared bond (Fig. \ref{Fig1}(a)) and a corner-shared bond (Fig. \ref{Fig1}(b))~\cite{Sugahara,Hirai}.
In CaIrO$_3$, the edge-shared bond is realized with an angle of 86$^{\circ}$~\cite{Sugahara,Hirai}.

Recently the resonance x-ray scattering showed that exchange interaction between 5d electrons is ferromagnetic (FM) in the case of the edge-shared bond (Fig. ~1(a)), while it is antiferromagnetic (AFM) in the case of the corner-shared bond~\cite{Ohgushi2}.
Therefore, it is very important and interesting to clarify the origin of the exchange interaction of the 5d electrons with very strong SO interaction. 
Jackeli and Khaliullin studied the mechanism of exchange interaction~\cite{Jackeli}.
They estimated an exchange interaction by projecting the superexchange spin-orbital models of $t_{2g}$ orbital with d$^5$ electrons onto the $\Gamma_7$ states.
They found that the exchange interaction is Ising FM called quantum compass model.
The origin of anisotropy of the exchange interaction was claimed to be the Hund's rule coupling.
However, the SO interaction was not considered explicitly in the perturbative process in ref.7.
As a result, the obtained exchange interaction was independent of SO interaction.
Furthermore, the physical mechanism of the appearance of anisotropic FM interaction was not clear. 

Generally speaking, the electronic state in the case of large SO coupling should be discussed on the basis of the J-J coupling scheme~\cite{Matsuura}.
However, the exchange interaction based on the J-J coupling has not been discussed so far irrespective of the long history of transition metal oxides.

In this letter, we study the exchange interaction between 5d$^5$ electrons in the t$_{2g}$ orbitals with large SO interaction on the basis of J-J coupling scheme as a typical model of 5d electron systems.
It is shown that the FM Ising interaction (quantum compass model) occurs in the edge-shared bond due to a selection rule of hybridization between $\Gamma_7$ and $\Gamma_8$ orbitals.
Simultaneously, an isotropic AFM Heisenberg interaction appears from the hybridization between $\Gamma_7$ orbitals.
These two interactions can be clearly distinguished by the second-order perturbation processes.
It is also found that the FM Ising interaction becomes very small in the large SO region, which is overlooked in ref.7.
As a result, we find that the interaction between 5d$^5$ electrons changes from the FM Ising to the AFM Heisenberg interaction as SO interaction increases.  

%%%%%%%%%%%%%%%%%%%%%%%%%%%%%%%%%%%%%%%%%%%%%%%%%%%%%%%%
%\section*{\it General Model}

\begin{figure}[h]
\begin{center}
\rotatebox{0}{\includegraphics[angle=-90,width=0.7 \linewidth]{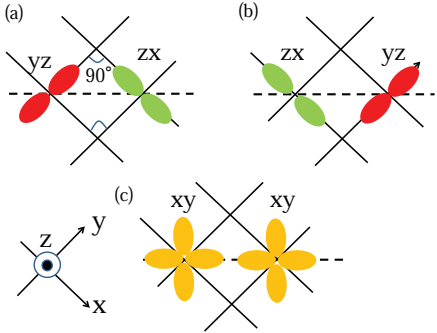}}
\caption{Schematic pictures of the relative positions of t$_{2g}$ orbitals on the nearest-neighbor Ir ions. The transfer integrals between d$_{yz}$ and d$_{zx}$ orbitals (a) and between d$_{zx}$ and d$_{yz}$ orbitals (b) are the same, and represented as $t$, while the transfer integral between d$_{xy}$ orbitals (c) is represented by $t_{xy}$ in eq.~(\ref{Hkin}).}
\label{Fig2}
\end{center}
\end{figure}

We consider the following model Hamiltonian for the edge-shared structure of CaIrO$_3$-type
\begin{eqnarray}
H = H_t + H_{{\rm int}} + H_{{\rm SO}}, \label{Ham}
\end{eqnarray}
where each term represents kinetic energy between the t$_{2g}$ orbitals, on-site Coulomb interaction, and on-site SO interaction for the t$_{2g}$ orbitals, respectively.
The kinetic energy is expressed as
\begin{equation}
H_t = \sum_{(i,j)\sigma} t (d^{\dagger}_{i,2\sigma} d_{j,3\sigma} + d^{\dagger}_{i,3\sigma} d_{j,2\sigma})
+ t_{xy} d^{\dagger}_{i,1\sigma}d_{j,1\sigma} + {{\rm h.c.}}, \label{Hkin} 
\end{equation}
where $\sum_{(i,j)}$ means the summation over the nearest-neighbor bonds, and $d_{i,\ell\sigma}$ is an annihilation operator of the $\ell$-th orbital with spin $\sigma$ on the $i$-th site ($d_{i,1\sigma}=d_{i,xy \sigma}$, $d_{i,2\sigma}=d_{i,yz\sigma}$, and $d_{i,3\sigma}=d_{i,zx\sigma}$). 
The relative positions of t$_{2g}$ orbitals on the nearest-neighbor Ir ions are shown in Fig.~\ref{Fig2}.
The parameters, $t$ and $t_{xy}$ represent the transfer integrals between $d_{yz}$ and $d_{zx}$ orbitals, and that between $d_{xy}$ orbitals, respectively.
In the actual material, the bond angle of Ir-O-Ir is 86$^{\circ}$ as shown in Fig.~\ref{Fig1}(a).
However, in order to obtain the essence of the material, we assume in this paper that the bond angle is 90$^{\circ}$ and neglect the degree of freedom of p orbitals on the oxygens.
This is the same approximation used in ref.\cite{Jackeli}.

For the Coulomb and SO interactions, we assume the following standard forms: 
\begin{equation}
\begin{split}
H_{{\rm int}} 
&= U_d \sum_{i} \sum_{\ell=1,2,3} n_{i,\ell\uparrow} n_{i,\ell\downarrow} \\
& + \frac{U_{d}^{\prime} - J_{d}}{2} \sum_{i,\sigma} \sum_{ \substack{ \ell,m=1,2,3 \\ (\ell \neq m) }} n_{i,\ell\sigma} n_{i,m\sigma} \\  
& + \frac{U_{d}^\prime}{2} \sum_{i, \sigma\neq \sigma^\prime} \sum_{\substack{ \ell,m=1,2,3 \\ (\ell \neq m) }} n_{i,\ell\sigma} n_{i,m\sigma^{\prime}} \\ 
& - \frac{J_{d}}{2} \sum_i \sum_{\substack{ \ell,m=1,2,3 \\ (\ell \neq m) }} ( d_{i,m\uparrow}^{\dag} d_{i,m\downarrow } d_{i,\ell\downarrow}^{\dag} d_{i,\ell\uparrow } \\
& \hspace{1cm} + d_{i,m\uparrow}^{\dag} d_{i,m\downarrow }^{\dagger} d_{i,\ell\uparrow} d_{i,\ell\downarrow } + {\rm h.c.}), 
\end{split} 
\end{equation} 
\begin{equation}
H_{{\rm SO}} = \frac{{\rm i}\zeta}{2} \sum_{i} \sum_{\substack{\ell mn = 1,2,3 \\ \sigma,\sigma^{\prime}}} \epsilon_{\ell mn} d_{i,\ell\sigma}^{\dagger} d_{i,m\sigma^{\prime}} \sigma_{\sigma\sigma^{\prime}}^{n},
\end{equation} 
where $n_{i,\ell\sigma} \equiv d_{i,\ell\sigma}^{\dagger} d_{i,\ell\sigma}$, $\epsilon_{lmn}$ is the Levi-Civita symbol, and $U_d$, $U_d^{\prime}$, $J_{d}$, and $\zeta$ are the intra- and inter-Coulomb interactions, FM exchange interaction (Hund's rule coupling), and the magnitude of SO interaction on $t_{2g}$ orbitals, respectively.  
These Coulomb interactions are expressed as
\begin{equation}
\begin{split}
U_d =& F_0 +4F_2 +36F_4 ,\\
U_d^{\prime} =& F_0 -2F_2 -4F_4,\\
J_d  =& 3F_2 +20F_4,
\end{split}
\end{equation}
where $F_0$, $F_2$, and $F_4$ are Slater-Condon parameters~\cite{Kamimura}.
They satisfy the relation, $U_{d}= U_{d}^{\prime} + 2J_d$.  

In order to clarify the mechanism of the exchange interaction in the J-J coupling scheme, we use $\Gamma_7$ and $\Gamma_8$ orbitals which diagonalize the SO interaction, $H_{\rm SO}$.
Four $\Gamma_8$ orbitals ($\phi_{1\alpha}$, $\phi_{1\beta}$, $\phi_{2\alpha}$, $\phi_{2\beta}$) are given by
\begin{equation}
\begin{split}
\phi_{1\alpha} =& \frac{1}{\sqrt{2}} (d_{yz\uparrow}   + {\rm i} d_{zx\uparrow} ),  \label{JJ1}\\ 
\phi_{1\beta}  =& \frac{1}{\sqrt{2}} (d_{yz\downarrow} - {\rm i} d_{zx\downarrow}), \\ 
\phi_{2\alpha} =& \frac{1}{\sqrt{6}} (2d_{xy\uparrow}  - d_{yz\downarrow} - {\rm i} d_{zx\downarrow}), \\ 
\phi_{2\beta}  =& \frac{1}{\sqrt{6}} (2d_{xy\downarrow} + d_{yz\uparrow} -{\rm i} d_{zx\uparrow}),
\end{split}
\end{equation}
and two $\Gamma_7$ orbitals ($\phi_{3\alpha}$, $\phi_{3\beta}$) are
\begin{equation}
\begin{split}
\phi_{3\alpha} =& \frac{1}{\sqrt{3}} (d_{xy\uparrow} + d_{yz\downarrow} + {\rm i} d_{zx\downarrow}), \label{JJ5}  \\
\phi_{3\beta}  =& \frac{1}{\sqrt{3}} (d_{xy\downarrow} - d_{yz\uparrow} + {\rm i} d_{zx\uparrow}). 
\end{split}
\end{equation}
The energy levels of $\Gamma_8$ and $\Gamma_7$ states are $\varepsilon_{\Gamma_8} = -\zeta/2$ and $\varepsilon_{\Gamma_7} = \zeta$, respectively.

In terms of $\phi$ orbitals, we rewrite the Coulomb interactions.
For example, when two electrons occupy $\phi_{1\alpha}$ and $\phi_{1\beta}$ orbitals, the wave function is given by
\begin{eqnarray}
\Phi({\rm r}_1, {\rm r}_2 ) = \frac{1}{\sqrt{2}}
 \left|
    \begin{array}{cc}
      \phi_{1\alpha}({\rm r}_1), & \phi_{1\alpha}({\rm r}_2)  \\
      \phi_{1\beta}({\rm r}_1), & \phi_{1\beta}({\rm r}_2) 
    \end{array}
  \right|.
\end{eqnarray}
Then the Coulomb interaction between $\phi_{1\alpha}$ and $\phi_{1\beta}$ orbitals are obtained as 
\begin{equation}
\begin{split}
U_{1} &\equiv \int\int d{\rm r}_1 d{\rm r}_2 \Phi({\rm r}_1, {\rm r}_2 )^{*}\frac{e^2}{|{\rm r}_1 - {\rm r}_2|}\Phi({\rm r}_1, {\rm r}_2 ) \\
      &= \langle -1, 1 ||-1,1 \rangle \\
      &= F_0 +F_2 +16F_4,
\end{split}
\end{equation}
where $\langle m_1, m_2 || m_1^{\prime},m_2^{\prime} \rangle$ is defined in ref. 8. 
The other Coulomb interactions between $\phi$ orbitals are given in the similar way. 
As a result, the effective Hamiltonian becomes  
\begin{equation}
\begin{split}
&H_{{\rm int}} 
= \sum_i \bigg[ U_1 (\hat{n}_{i,1\alpha} \hat{n}_{i,1\beta} + \hat{n}_{i,2\alpha} \hat{n}_{i,2\beta}) + U_2 \hat{n}_{i,3\alpha}\hat{n}_{i,3\beta} \\
&+ U_1^{\prime} \sum_{k,k^{\prime}=\alpha,\beta} \hat{n}_{i,1k} \hat{n}_{i,2k^{\prime}} \\   
&+ U_2^{\prime} (\hat{n}_{i,1\alpha} \hat{n}_{i,3\alpha} + \hat{n}_{i,1\beta}  \hat{n}_{i,3\beta}) \\  
&+ U_3^{\prime} (\hat{n}_{i,1\beta}  \hat{n}_{i,3\alpha} + \hat{n}_{i,1\alpha} \hat{n}_{i,3\beta}) \\
&+ U_4^{\prime} (\hat{n}_{i,2\alpha} \hat{n}_{i,3\alpha} + \hat{n}_{i,2\beta}  \hat{n}_{i,3\beta}) \\ 
&+ U_5^{\prime} (\hat{n}_{i,2\beta}  \hat{n}_{i,3\alpha} + \hat{n}_{i,2\alpha} \hat{n}_{i,3\beta}) \\ 
&+ J_1 (
\hat{\phi}_{i,1\alpha}^{\dagger} \hat{\phi}_{i,1\beta}^{\dagger} \hat{\phi}_{i,2\alpha} \hat{\phi}_{i,2\beta} 
+ \hat{\phi}_{i,1\alpha}^{\dagger} \hat{\phi}_{i,1\beta}^{\dagger} \hat{\phi}_{i,3\alpha} \hat{\phi}_{i,3\beta}
+ {\rm h.c.} )\\ 
&+ J_2 (\hat{\phi}_{i,2\alpha}^{\dagger} \hat{\phi}_{i,2\beta}^{\dagger} \hat{\phi}_{i,3\alpha} \hat{\phi}_{i,3\beta} + {\rm h.c.})\\
&+ J_3 (\hat{\phi}_{i,1\beta}^{\dagger} \hat{\phi}_{i,3\alpha}^{\dagger} \hat{\phi}_{i,2\beta} \hat{\phi}_{i,3\beta} + \hat{\phi}_{i,2\alpha}^{\dagger} \hat{\phi}_{i,3\alpha}^{\dagger} \hat{\phi}_{i,1\alpha} \hat{\phi}_{i,3\beta}+ {\rm h.c.})\\
&+ J_4 (\hat{\phi}_{i,2\beta}^{\dagger} \hat{\phi}_{i,3\alpha}^{\dagger} \hat{\phi}_{i,2\alpha} \hat{\phi}_{i,3\beta} + {\rm h.c.}) \bigg],
\end{split}
\end{equation}
with
\begin{equation}
\begin{split}
U_1 &= F_0 +F_2 +16F_4, \hspace{1cm}
U_2 = F_0  +\frac{28}{3}F_4, \\
U_1^{\prime} &=  F_0 -3F_2 -\frac{32}{3}F_4, \hspace{1cm}
U_2^{\prime} =  F_0 -F_2 +\frac{8}{3}F_4, \\
U_3^{\prime} &=  F_0 -4F_2 -\frac{52}{3}F_4, \hspace{1cm}
U_4^{\prime} =  F_0 -2F_2 -4F_4, \\
U_5^{\prime} &= U_1^{\prime} = F_0 -3F_2 -\frac{32}{3}F_4, \\
J_1  &= 4F_2 +\frac{80}{3}F_4, \hspace{1cm} 
J_2  = 5F_2 +\frac{350}{9}F_4, \\
J_3  &= \sqrt{3}(F_2 +\frac{20}{3}F_4), \hspace{1cm}
J_4  = -2(F_2 +\frac{20}{3}F_4),
\end{split}
\end{equation}
where $\hat{\phi}_{i,\gamma k}$ ($\gamma=1,2,3, k=\alpha, \beta$) is an annihilation operator of the $\phi_{\gamma k}$ orbital (eqs.($\ref{JJ1}$) - ($\ref{JJ5}$)) on the $i$-th site, and $\hat{n}_{i,\gamma k}=\hat{\phi}_{i,\gamma k}^{\dagger}\hat{\phi}_{i,\gamma k}$.

In the similar way, we rewrite the kinetic energy in terms of $\hat{\phi}_{i,\gamma k}$ to obtain
\begin{equation}
\begin{split}
H_t &= t_{xy} \sum_{(i,j),k=\alpha,\beta} \bigg[ 
  \frac{1}{3}        \hat{\phi}_{i,3k}^{\dagger} \hat{\phi}_{j,3k} 
+ \frac{\sqrt{2}}{3} \hat{\phi}_{i,3k}^{\dagger} \hat{\phi}_{j,2k} \\ & \hspace{2.5cm} 
+ \frac{2}{3}        \hat{\phi}_{i,2k}^{\dagger} \hat{\phi}_{j,2k} +{{\rm h.c.}}  \bigg] \\
&+ t \sum_{(i,j)} \sum_{\substack{k,k^{\prime}=\alpha,\beta \\ (k \neq k^{\prime})}}
\bigg[ -\frac{2{\rm i}}{\sqrt{6}} \hat{\phi}_{i,3k}^{\dagger} \hat{\phi}_{j,1k^{\prime}} \\
& \hspace{2.5cm} + \frac{{\rm i}}{\sqrt{3}} \hat{\phi}_{i,2k}^{\dagger} \hat{\phi}_{j,1k^{\prime}} 
+{{\rm h.c.}} \bigg]. 
\end{split}
\end{equation}
Note that the hopping matrix elements are characteristic.
For example, an electron in $\phi_{1\alpha}$ orbital can hop only to $\phi_{3\beta}$ and $\phi_{2\beta}$.

\begin{figure}[h]
\begin{center}
\rotatebox{0}{\includegraphics[angle=-90,width=0.8 \linewidth]{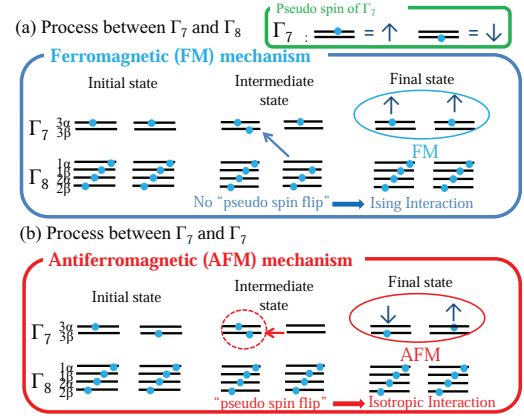}}
\caption{Mechanism of (a) ferromagnetic (FM) interaction and (b) antiferromagnetic (AFM) interaction. The top of this figure shows the definition of pseudo-spin in $\Gamma_7$ orbital.}
\label{Fig3}
\end{center}
\end{figure}

Using these $\Gamma_7$ and $\Gamma_8$ orbitals, we can describe the ground state of the on-site Hamiltonian, $H_{\rm int} + H_{\rm SO}$.
Since Ir ion has five electrons, the four electrons occupy $\Gamma_8$ orbitals, and the remaining one electron occupies one of the $\Gamma_7$ orbitals, i.e., $\phi_{3\alpha}$ or $\phi_{3\beta}$.
These two states have the degenerate energy $\epsilon_{\Gamma_7}=\frac{3}{2}\zeta$.
We regard these states as pseudo-spin states, i.e., the state in which one electron occupies $\Gamma_{3\alpha}$($\Gamma_{3\beta}$) is regarded as  up(down) pseudo-spin state (see the top panel of Fig. \ref{Fig3}).
In the following, we study the exchange interactions between pseudo-spins in the second order perturbation with respect to $H_t$.  

Figure \ref{Fig3}(a) shows one of the second order processes.
In the initial state, one electron occupies $\Gamma_7$ orbital and four electrons occupy $\Gamma_8$ orbitals on each Ir ions.
In the intermediate state, one electron moves from one of $\Gamma_8$ orbitals to $\phi_{3\beta}$ of the nearest neighbor Ir ion as shown in the middle panel of Fig. \ref{Fig3}(a).
In the final state, the pseudo-spin flip does not occur in this process because each $\Gamma_8$ orbital hybridizes only one of the $\Gamma_7$ orbitals as shown in eq. (12). 
Therefore, the effective exchange interaction between pseudo-spins in this process is Ising type.  

Using this second order process, we obtain the energy of $\uparrow\uparrow$ ($\uparrow\downarrow$) state $E_{\uparrow\uparrow}$ ($E_{\uparrow\downarrow}$) as 
\begin{equation}
\begin{split}
E_{\uparrow\uparrow} 
&= - \frac{\frac{4}{9}t_{xy}^2}{E_0 +\frac{3}{2}\zeta}  - \frac{\frac{4}{3}t^2}{E_1 + \frac{3}{2}\zeta}, \\  
E_{\uparrow\downarrow} 
&= - \frac{\frac{4}{9}t_{xy}^2}{E_2 +\frac{3}{2}\zeta}  - \frac{\frac{4}{3}t^2}{E_3 + \frac{3}{2}\zeta},
\end{split}
\end{equation} 
with $E_0 = U_d^{\prime}$, $E_1 = U_d^{\prime} -\frac{2}{3}J_d$, $E_2 = U_d^{\prime} -\frac{1}{3}J_d$, and $E_3 = U_d^{\prime} +\frac{1}{3}J_d$, respectively.
The denominator is the energy difference between the intermediate and initial state. 
Since $E_{\uparrow\uparrow} < E_{\uparrow\downarrow}$, the exchange interaction can be expressed as FM Ising interaction, $-2J_{\rm FM}S_{iz}S_{jz}$, where $J_{\rm FM}=E_{\uparrow\downarrow}-E_{\uparrow\uparrow}$ and ${\bf S}_{i}$ represent the pseudo-spin operator on the $i$-th site.
In the limit of large SO interaction $\zeta$, (i.e., $\zeta \gg F_0$, $F_2$ and $F_4$), we obtain  
\begin{equation}
J_{\rm FM} \simeq \frac{16}{81}\bigr[ t_{xy}^2 +3t^2 \bigr] \frac{J_d}{\zeta^2}. \label{FM}
\end{equation}

Figure \ref{Fig3}(b) shows another second order process.
In the intermediate state, one electron moves from $\Gamma_7$ orbital to $\Gamma_7$ orbital of the nearest-neighbor Ir ion as shown in the middle panel of Fig. \ref{Fig3}(b). 
This process is same as the superexchange process discussed by Anderson~\cite{Anderson}, and a pseudo-spin flip process exists.
Therefore, this exchange interaction is AFM Heisenberg interaction written as $J_{\rm AF}{\bf S}_{i}\cdot{\bf S}_{j}$ with
\begin{eqnarray}
J_{\rm AF} = \frac{\frac{4}{9}t_{xy}^{2}}{U_2}.
\end{eqnarray}
Note that $J_{\rm AF}$ is independent of the SO interaction.

Adding these two processes, the effective Hamiltonian between the two pseudo-spins on the nearest-neighbor site is given by
\begin{eqnarray}
H_{\rm eff} = -2J_{\rm FM}S_{1z}S_{2z} +J_{\rm AF}{\bf S}_{1} \cdot {\bf S}_{2}.  \label{KHM}
\end{eqnarray}
The first term on the right-hand side is same as quantum compass-Heisenberg interaction.
In the present calculation, d$_{yz}$ and d$_{zx}$ orbitals are not equivalent to d$_{xy}$ orbital, because z-axis is special as shown in Fig. \ref{Fig2}.
Thus, the Ising interaction, $J_{\rm FM}S_{1z}S_{2z}$, occurs.
If we change the axis as $(x,y,z) \rightarrow (y,z,x)$, x-axis becomes a special axis, and $J_{\rm FM}S_{1x}S_{2x}$ occurs.
This kind of interaction which depends on the direction is called quantum compass model.

Note that $J_1 \sim J_4$ terms in eq. (10) have been neglected in the intermediate states of the second order perturbation.
If these terms are considered, an additional term, $H_{\rm add} = J_{\rm add}(S_{1+}S_{2+}+S_{1-}S_{2-})$, is obtained with $J_{\rm add} \propto tt_{xy}J_d/\zeta^2$~\cite{Rau,Yamaji,Nasu} which does not conserve $S_z$.
In this paper, we focus on $J_{\rm FM}$ and $J_{\rm AFM}$, and
$J_{\rm add}$ will be discussed in the forthcoming paper by comparing with the numerical diagonalization.    

\begin{figure}[h]
\begin{center}
\rotatebox{0}{\includegraphics[angle=0,width=0.8 \linewidth]{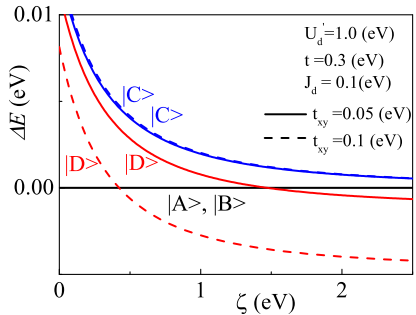}}
\caption{Eigenvalues measured from $E_A$ as a function of the SO interaction, $\zeta$, for $t_{xy}=0.05$eV (solid line) and $t_{xy}=0.1$eV (dashed line). 
The parameters, $U_d^{\prime}=1.0$eV, $t=0.3$eV, and $J_d=0.1$eV are chosen.}
\label{Fig4}
\end{center}
\end{figure}

The eigenstates of the two-site quantum compass-Heisenberg model (eq.(\ref{KHM})) are 
$|A \rangle \equiv \uparrow \uparrow$,
$|B \rangle \equiv \downarrow \downarrow$,
$|C \rangle \equiv \frac{1}{\sqrt{2}}\biggr[ \uparrow \downarrow  + \downarrow \uparrow  \biggr]$, and $|D \rangle \equiv \frac{1}{\sqrt{2}}\biggr[ \uparrow \downarrow  - \downarrow \uparrow \biggr]$.
Figure \ref{Fig4} shows the eigenvalues measured from $E_A$ as a function of the SO interaction for $t_{xy}=0.05$eV (solid line) and $t_{xy}=0.1$eV (dashed line).
The parameters $U_d^{\prime}=1.0$eV, $J_d=0.1$eV, and $t=0.3$eV are assumed.  

The state $|B\rangle$ is always degenerate with $|A\rangle$.
For small values of $\zeta$, the ground state is $|A\rangle$ and $|B\rangle$, since the FM Ising interaction is dominant.
On the other hand, when $\zeta$ is large, the FM interaction becomes small (see eq.(\ref{FM})) and the ground state becomes a pseudo-spin singlet $|D\rangle$.
The energy difference between $E_D$ and $E_A$ is of the order of 0.001eV which is consistent with the ab-initio calculation~\cite{Bogdanov}.

Finally, we comment on the exchange interaction in the case of the corner-shared bond shown in Fig.\ref{Fig1}(b).
In a similar way, we obtain the exchange interaction with quantum compass and Heisenberg interactions such as eq. (\ref{KHM}).
However, we find that $J_{\rm FM} \ll J_{\rm AF}$ in the realistic parameter region. 
Thus, we can regard the exchange interaction of the corner shared bond as the conventional AFM Heisenberg interaction.     

We discuss the difference between our results and ref.7. 
In ref.7, the exchange interaction was derived by projecting the superexchange model, which is introduced for $\zeta =0$, onto the $\Gamma_7$ state.
As a result, the exchange interaction was independent of $\zeta$. 
However, as shown in the present paper, the FM Ising interaction is strongly on $\zeta$.

\begin{figure}[h]
\begin{center}
\rotatebox{0}{\includegraphics[angle=-90,width=0.8 \linewidth]{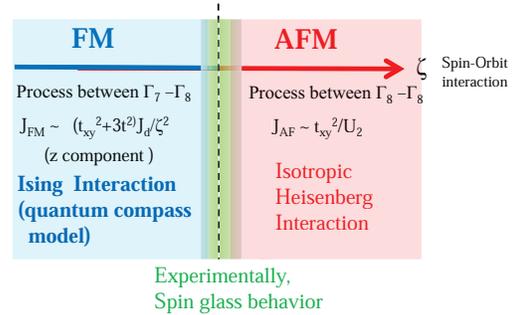}}
\caption{Schematic phase diagram of the exchange interaction. 
Blue and red lines indicate the region of ferromagnetic (FM) and antiferromagnetic (AFM) interaction, respectively.
The green region shows the spin glass region expected in the real materials.  }
\label{Fig5}
\end{center}
\end{figure}

In this paper, we clarified the mechanism of the exchange interaction between 5d electrons on t$_{2g}$ orbitals based on the J-J coupling scheme. 
The results are summarized in Fig.\ref{Fig5}.
The second-order perturbation shows that the FM Ising interaction occurs due to the hybridization between the $\Gamma_7$ and $\Gamma_8$ orbitals in which no pseudo-spin flip process occurs.
As SO interaction increases, the exchange interaction changes from the FM Ising  (quantum compass) to the isotropic AFM Heisenberg interaction.
Due to the defects and impurities in the real materials, a spin glass behavior is expected around the edge of the FM and AFM states.

The authors acknowledge H. Nakada and Y. Fukusumi who have discussed as a fulfillment of the degree of bachelor in the department of physics, the university of Tokyo.
One of the authors (H.M.) acknowledge helpful discussions with J. Nasu, Y. Yamaji, and J. Kishine. 
This work is supported by Grant-in-Aid for Scientific Research on Innovative Areas gUltra Slow Muon Microscopeh (No. 23108004) from the@Ministry of Education, Culture, Sports, Science and Technology, Japan.
We are supported by Grant-in-Aid for Scientific Research from the Japan Society for the Promotion of Science (No. 25220803), and (No. 24244053).

\end{document}